\documentclass[12pt]{article}
\usepackage{amsmath,amssymb}    
\usepackage{amsfonts}
\usepackage{enumerate}
\usepackage{ytableau}
\usepackage{hyperref}
\usepackage{bm}
\usepackage{tikz}
\usepackage{blkarray}  
\begin{document}
\def\be{\begin{equation}}
\def\bea{\begin{eqnarray}}
\def\ee{\end{equation}}
\def\eea{\end{eqnarray}}
\def\d{\partial}
\def\eps{\varepsilon}
\def\la{\lambda}
\def\b{\bar}
\def\nn{\nonumber \\}
\def\p{\partial}
\def\t{\tilde}
\def\h{{1\over 2}}
\def\cp{\mathbb{CP}^1}
\makeatletter
\def\blfootnote{\xdef\@thefnmark{}\@footnotetext}  
\makeatother

\title{\textbf{$\lambda$-deformed $AdS_5\times S^5$ superstring \vspace{0.15cm} from 4D Chern-Simons theory}}

\vspace{14mm}
\author{
Jia Tian$^{1,3}$, Yi-Jun He$^{1}$,and Bin Chen$^{1,2,3}$\footnote{wukongjiaozi, yjhe96, bchen01@pku.edu.cn}
	}

\date{}

\maketitle

\begin{center}
	{\it
		$^{1}$School of Physics and State Key Laboratory of Nuclear Physics and Technology,\\Peking University, No.5 Yiheyuan Rd, Beijing 100871, P.~R.~China\\
		\vspace{2mm}
		$^{2}$Collaborative Innovation Center of Quantum Matter, No.5 Yiheyuan Rd, Beijing 100871, P.~R.~China\\
		$^{3}$Center for High Energy Physics, Peking University, No.5 Yiheyuan Rd, Beijing 100871, P.~R.~China
	}
	\vspace{10mm}

\end{center}

\begin{abstract}
We present the construction of the $\lambda$-deformation of $AdS_5\times S^5$ superstring from the four dimensional Chern-Simons-type gauge theory. The procedure  is applicable to all the semi-symmetric coset models and generalizes the previous construction of $\lambda$-deformation of the symmetric coset models.
\end{abstract}

\baselineskip 18pt

\newpage
\tableofcontents

\section{Introduction}
The classical integrable systems can be described in the Lax formalism. Therefore in some sense classifying the classical integrable systems is equivalent to classifying Lax connections. For classical  Hamiltonian systems there does exist a general procedure, due to Zakharov and Shabat, to construct integrable systems from the Lax pairs \cite{ZS}. In this so-called Zakharov-Shabat construction, the Lax pairs are characterized by their analytic properties and the underlying semi-simple Lie algebras\footnote{For details, see the classic book \cite{YellowBook}.}. There is also a field theory analogue of Zakharov-Shabat construction \cite{ZSfield} by extending the semi-simple Lie algebra to an infinite dimensional algebra such as the central-extended loop algebra or affine Kac-Moody algebra. But this construction has not been carefully considered or fully explored. The recently developed affine Gaudin model approach is closest to such a construction \cite{Gaudin}. The affine Gaudin model is  in the Hamiltonian formulation while sometimes it is more convenient to describe classical field theories in the Lagrangian formulation. 

In the past few years, a new approach to study integrable models from the point of view of four dimensional Chern-Simons-type (4D CS) gauge theory \cite{Costello:2013zra} has been developed in a series of papers \cite{Costello:2017dso,Costello:2018gyb,CY3}. In particular,  Costello and Yamazaki \cite{CY3} proposed a 
Lagrangian approach  to construct two dimensional integrable field theories (2D IFTs) from a 4D CS gauge theory with a meromorphic one-form $\omega$. It turns out that the 4D approach is closely related to the affine Gaudin model approach \cite{Relation}. From the 4D CS point of view, the resulted 2D IFTs are completely characterized by the choice of the one-form $\omega$ and the boundary conditions of the gauge fields. The construction has shown its power by covering a wide variety of 2D IFTs, including principle chiral models (PCM), Wess-Zumino-Witten models, sigma models whose target space are symmetric cosets, the superstring on AdS$_5 \times $ S$^5$, etc.\cite{CY3}. It was conjectured that the all known 2D IFTs could arise from such a construction. Following the work in \cite{CY3}, many interesting 2D IFTs with deformations, including the Yang-Baxter deformation \cite{BosonicYangBaxter,DelducEta, Yoshida} and $\lambda$-deformation \cite{Lambda,LambdaCoset},    have been realized in this approach \cite{JT,Uni, COUPLED,HOLOLAMBDA,YB1,YB2,STRING}.

In this paper our goal is to realize the $\lambda$-deformation of the $AdS_5\times S^5$  superstring or other semi-symmetric coset models\footnote{The homogeneous Yang-Baxter deformation of $AdS_5\times S^5$  superstring was considered in \cite{YB2}.}. The strategy comes from the observations that in the 4D approach the $\lambda$-deformation can be performed by splitting the double poles of the one-form $\omega$ into a pair of simple poles, and the symmetric coset models can be constructed by introducing the covering space \cite{CY3,JT}.

The paper has the following organization. In section \ref{Review} we briefly describe the 4D CS approach. In section \ref{lambda} and section \ref{Superstring} we show the constructions of $\lambda$-deformation of Principle Chiral Model (PCM) and $AdS_5\times S_5$ superstring which are two main ingredients of our following construction. The section \ref{LambdaSuperstring} is devoted to the derivation of the $\lambda$-deformation of $AdS_5\times S_5$ superstring in the pure spinor formulation and in the Appendix, we give the construction in Green-Schwarz formulation.

\section{2D IFTs from 4D CS}
\label{Review}
\renewcommand{\theequation}{2.\arabic{equation}}
\setcounter{equation}{0}
In this section we briefly describe how to derive 2D integrable sigma models from 4D Chern-Simons-type gauge theory.  We begin with the action of the 4D Chern-Simons-type gauge theory \footnote{Here we follow the convention in \cite{Uni}.} 
\bea \label{4Daction} 
S[A]=\frac{i}{4\pi} \int_{\mathbb{R}^{2} \times \mathbb{CP}^1} \omega \wedge CS(A),
\eea 
where 
\bea 
&&A=A_\sigma d\sigma+A_\tau d\tau+A_{\bar{z}}d\bar{z},\quad \omega=\omega(z)dz,\\
&&CS(A)=\langle A,dA+\frac{2}{3}A\wedge A\rangle.
\eea 
Here $\omega(z)$ is a meromorphic function on $\cp$ and it can be identified with the twist function which also plays the crucial role in the construction of integrable field theories from the affine Gaudin model approach \cite{Gaudin}. Varying the action \eqref{4Daction} with respect to the gauge field $A$ gives
the bulk equation of motion
\bea \label{BulkEOM}
\omega\wedge F(A)=0,\quad F(A)=dA+A\wedge A=0,
\eea 
and the boundary equation of motion
\bea \label{BEOM}
d\omega\wedge \langle A,\delta A\rangle=0.
\eea 
The bulk equation of motion is just the flatness condition of the gauge field. To study the boundary equation of motion it is useful to rewrite \eqref{BEOM} in terms of the coordinates: 
\bea \label{Boundary}
\sum_{x\in \mathfrak{p}}\sum_{p\geq 0}^{}(\text{Res}_x~ \xi^p_x \omega)\epsilon_{ij}\frac{1}{p!} \p^p_{\xi_x}\langle A_i, \delta A_j \rangle |_x=0,
\eea 
where  $\mathfrak{p}$ is the set of poles of the one-form $\omega$, $\xi_x$ is a local holomorphic coordinate around $x \in \mathfrak{p}$ and $i,j= \sigma,\tau$ are the coordinates of $\mathbb{R}^{2}$. In our construction we will only encounter  $\omega$ with simple poles and double poles. For the double poles we always choose the Dirichlet boundary condition, i.e.
\bea \label{Dirichlet}
A=\delta A=0.
\eea 
Considering only the simple poles, the boundary equation of motion \eqref{BEOM} simplifies to
\bea \label{BSimple}
\sum_{x\in \mathfrak{p}}(\text{Res}_x \omega)\epsilon_{ij}\langle A_i ,\delta A_j\rangle |_x=0.
\eea  
The zeros of $\omega$  constrain the gauge fields as well. To have a non-degenerate propagator the gauge fields must have the poles at the positions of the zeros of $\omega$.  This could be induced by introducing  some defect operators\footnote{These are called the disorder defects in \cite{CY3}.}  inserted at those positions. Then the 2D IFT is totally determined by the boundary conditions and the defect operators.

The Lax connection of the 2D IFT is directly related to the 4D gauge field through a gauge transformation
\bea \label{ConnectionLax}
A=-d\hat{g} \hat{g}^{-1} +\hat{g}L\hat{g}^{-1},
\eea 
for some regular $\hat{g}:\mathbb{R}^{2}\times \cp \rightarrow G$  such that $L_{\bar{z}}=0$. In terms of the Lax connection the bulk equation of motion \eqref{BulkEOM} reads \footnote{The light-cone coordinates are defined as $\sigma_\pm=\frac{1}{2}(\tau\pm i\sigma)$. }
\bea 
[\p_++L_+,\p_-+L_-]=0,\quad \omega \wedge \p_{\bar{z}}L(z,\tau,\sigma)=0.
\eea
The second identity implies that the positions of the poles of $L$ coincide with the positions of the zeros of $\omega$.  When the 4D field $\hat{g}$ satisfies the archipelago conditions introduced in \cite{Uni} we can localize the four dimensional field  to a two dimensional  field $g_x$ such that the 4D action \eqref{4Daction} is reduced to a 2D action \cite{Uni}
\bea \label{Action}
S[\{g_x\}_{x\in \mathfrak{p}}]=\frac{1}{2} \sum_{x\in\mathfrak{p}} \int_{\mathbb{R}^{2}} \langle \text{Res}_x \omega \wedge L,g_x^{-1}dg_x\rangle-\frac{1}{2}\sum_{x\in \mathfrak{p}}(\text{Res}_x \omega) I_{WZ}[g_x].
\eea 
This resulted 2D theory has two kinds of gauge redundancies. Firstly there is always an overall gauge transformation $g_x \rightarrow  g_x h, h \in G$ which reflects the redundancy in the definition of $g_x$ in terms of $A$.  Secondly if the gauge field $A$ at $x\in \mathfrak{p}$ does not vanish but takes values in some algebra $\mathfrak{h}$ whose corresponding group is $H$ then we can perform a gauge transformation $g_x \rightarrow u_x g_x, u_x\in H$.
\subsection{Reality condition}
In the 4D CS theory the gauge fields generally take values in a complex Lie algebra $\mathfrak{g}^{\mathbb{C}}$. So in order to construct a physical 2D model, the gauge field and the 1-form $\omega$ are required to satisfy proper reality conditions \cite{Uni}. Let $\theta:\mathfrak{g}^{\mathbb{C}}\rightarrow \mathfrak{g}^{\mathbb{C}}$ be an anti-linear involutive automorphism of the complex Lie algebra such that
\bea 
\overline{\langle B,C\rangle}=\langle \theta(B),\theta(C)\rangle,\quad \forall B,C \in \mathfrak{g}^{\mathbb{C}}.
\eea 
The involution automorphism can be lifted an involution automorphism of the group $\Theta: G^{\mathbb{C}}\rightarrow G^{\mathbb{C}} $. A real form $\mathfrak{g}$ of $\mathfrak{g}^{\mathbb{C}}$ is defined as the subset of the fixed point under $\theta$. Recall the complex conjugation of $z$ on $\mathbb{CP}^1$ is defined as $C:z\rightarrow \bar{z}$ so we should impose the following reality condition on the gauge field
\bea \label{real2}
C^{*}A=\theta(A),\quad A \in \mathfrak{g}^{\mathbb{C}}
\eea 
and the gauge transformation should also satisfy
\bea 
C^{*}g=\Theta(g),\quad g \in G^{\mathbb{C}}.
\eea 
Then the reality condition of the action \eqref{4Daction} is ensured by the condition \cite{Uni}
\bea \label{real1}
C^{\star}\omega=\overline{\omega}.
\eea

To conclude this section, let us summarize the general procedures to construct a 2D IFT:
\begin{enumerate}
	\item Choose a meromorphic one-form $\omega=\omega(z)dz$.
	\item Specify the boundary conditions at the positions of the poles of $\omega$.
	\item Remove the gauge redundancies \footnote{One can also choose to remove the gauge redundancies in the end.} in $g_x,x\in \mathfrak{p}$.
	\item Make an ansatz of the Lax connection with the poles at the positions of  the zeros of $\omega$.
	\item Solve the Lax connection by substituting \eqref{ConnectionLax} into the boundary conditions.
	\item Substitute the Lax connection into \eqref{Action} to get the action of the 2D IFT.
\end{enumerate}

\section{$\lambda$-deformation of PCM}
\label{lambda}
\renewcommand{\theequation}{3.\arabic{equation}}
\setcounter{equation}{0}
The prototypical example is the construction of the principle chiral model (PCM) \cite{Uni}. The corresponding one-form has two double poles and two simple zeros. To construct the $\lambda$-deformed PCM, one can split one double pole into two simple poles. As a result the one-form can be chosen to be \cite{Uni}
\bea \label{lambdaOneform}
\omega=\frac{1}{1-\eta^2}\frac{1-z^2}{z^2-\eta^2}dz.
\eea  
The boundary condition \eqref{BSimple} which needs to be satisfied is 
\bea 
\frac{1}{2\eta}\epsilon_{ij}(\langle A_i ,\delta A_j\rangle |_\eta-\langle A_i ,\delta A_j\rangle |_{-\eta})=0.
\eea 
The solution which leads to the $\lambda$-deformation  is  requiring $(A|_\eta,A|_{-\eta})\in \mathfrak{g}^\delta$ to take values in a Lagrangian subalgebra of $\mathfrak{g}\oplus \mathfrak{g}\equiv \mathfrak{d}$ as\footnote{Actually we have an Manin triple $\left(\mathfrak{d}, \mathfrak{g}_{R}, \mathfrak{g}^{\delta}\right)$ for some solutions $R$ of modified classical Yang-Baxter equation and $\mathfrak{g}_{R}:=\{((R-1){x},(R+1){x}|{x}\in\mathfrak{g}\}$. The other Manin pair $(\mathfrak{g}_R,\mathfrak{d})$ will lead to Yang-Baxter model \cite{Uni}. } 
\bea \label{BoundaryLambda}
\mathfrak{g}^{\delta}=\{(x,x)|x\in\mathfrak{g}\}\,\quad (\mathfrak{g}^\delta,\mathfrak{d} ) \text{ is a Manin pair.}
\eea 
In other words, we will identify the gauge fields at the two boundaries $z=\pm \eta$. Since $z=\infty$ is a double pole, we will use the Dirichlet boundary condition. To remove the gauge redundancy we first use the overall gauge symmetry to set $g_\infty=I$ then use the local gauge symmetry $H=\{(h,h)|h\in G\}$ to fix $(g_\eta,g_{-\eta})=(g,1)$.
Because the one-form \eqref{lambdaOneform} has zeros at $\pm 1$ and the gauge field vanishes at $z=\infty$ so the ansatz of the Lax connection can be 
\bea \label{ansatzLambda}
\mathcal{L}_+=\frac{V_+}{z-1},\quad \mathcal{L}_-=\frac{V_-}{z+1},
\eea 
where $V_\pm$ are regular functions. Rewriting the gauge fields in terms of the Lax connection through \eqref{ConnectionLax} and substituting into the boundary condition give a set of equations of $V_\pm$. The solution is
\bea 
&&V_+=(\eta-1)(\operatorname{Ad}_g+\frac{\eta-1}{\eta+1})^{-1}\operatorname{Ad}_g j_+,\\ &&V_-=(\eta+1)(\operatorname{Ad}_g+\frac{\eta+1}{\eta-1})^{-1}\operatorname{Ad}_g j_-,
\eea 
where $\operatorname{Ad}_g x=g x g^{-1}$ is the adjoint conjugation. Introducing the parameters $\lambda$ and $k$ as
\bea\label{Para}
\lambda=\frac{1+\eta}{1-\eta}, \quad k=-1/(4\eta),
\eea
and substituting the Lax connection into \eqref{Action} leads to the action of $\lambda$--deformed PCM \cite{Lambda}
\bea 
&&\mathcal{S}[g]=k \int d\sigma_+\wedge d\sigma_- \left[\langle j_+,j_-\rangle +2 \langle (\lambda^{-1}-\operatorname{Ad}_g)^{-1}\operatorname{Ad}_g j_+,j_-\rangle\right] +k I_{WZ}(g).
\eea 

To conclude this section, let us make some comments about the construction. 
\begin{itemize}
	\item In \eqref{Para} it seems that $\lambda$ and $k$ are related to each other, however we have  freedom to put a prefactor in the one-form such that these parameters become independent.
	\item In the ansatz of the Lax connection \eqref{ansatzLambda}, we intentionally distribute the two poles into the two components of the Lax connection to avoid the appearance of the double poles in the flatness condition. In principle the most general ansatz should be
	\bea 
	\mathcal{L_\pm}=\frac{V_1^\pm}{z-1}+\frac{V_{-1}^{\pm}}{z+1}.
	\eea 
	\item Because we have assumed the algebra $\mathfrak{g}$ to be real in the beginning so the resulted 2D IFT is  real. Otherwise one has to consider additional reality conditions \cite{Uni}.
	\item To get a non-trivial 2D IFT, the one-form must at least have two poles, otherwise one can always use the overall gauge symmetry to trivialize the field $g=I$. 
	\item To construct the symmetric coset  model, one can start with the PCM and  modulo a $\mathbb{Z}_2$ discrete symmetry. This quotient will not identify the fields at the two boundaries instead the quotient relates the two fields through a $\mathbb{Z}_2$ involution as $g_2=\rho (g_1),\rho^2=1$. To construct the $\lambda$ deformation of the coset model, one can split both of the double poles into two pairs of simple poles and apply the boundary condition \eqref{BoundaryLambda} \cite{JT}.
	\item One 2D IFT can be constructed from different set-ups of the 4D CS theory. For example, the (Yang-Baxter deformation or $\lambda$-deformation of) symmetric coset model has been obtained in \cite{YB1} and in \cite{JT} with different settings.
\end{itemize}

\section{The $AdS_5\times S^5$ superstring}
\label{Superstring}
\renewcommand{\theequation}{4.\arabic{equation}}
\setcounter{equation}{0}
Let us start with the undeformed theory. In the original paper \cite{CY3}, the $AdS_5\times S^5$ superstring was constructed as a special example of the generalized Riemannian symmetric spaces. Recently it was constructed from a different setting of the 4D CS theory \cite{YB2}. Interestingly the authors of \cite{YB2} found that by choosing different boundary conditions the same one-form can lead to either the $AdS_5\times S^5$ superstring or homogeneous Yang-Baxter deformation of the superstring. To construct the $\lambda$-deformation of the superstring, we find that starting with original setting in \cite{CY3} is more convenient.

The semi-symmetric coset space $AdS_5\times S^5$ is associated with a $\mathbb{Z}_4$ graded algebra $\mathfrak{psu}(2,2|4)=\mathfrak{g}=\mathfrak{g}^{(0)}\oplus \mathfrak{g}^{(1)}\oplus \mathfrak{g}^{(2)}\oplus \mathfrak{g}^{(3)}$. We will extend $\mathfrak{g}$ to $\mathfrak{g}^{\mathbb{C}}$ so that the 4D gauge field will  in general take values in $\mathfrak{su}(2,2|4)^{\mathbb{C}}$. The construction is also suitable for other semi-symmetric coset spaces with $\mathbb{Z}_4$ grading, for example  $AdS_3\times S^3$.
There exists an  automorphism $\rho$ of order 4 for the lie algebra $\mathfrak{g}$. 
Starting with a Riemann surface equipped with a holomorphic 1-form $dz$,  we introduce a cut of degree $4$ at  $[0,1]$ and when the gauge fields cross this cut we apply the automorphism $\rho$. Then we go to the 4-fold covering space of the $z$--plane by introducing the new coordinate $u$ through
\bea 
z=-\frac{1}{u^4-1}.
\eea 
In the 4-fold covering space, the one-form $dz$ pulls back to\footnote{This is also the one-form considered in \cite{YB2}.}
\bea \label{SuperOneForm}
\omega=\frac{4u^3 du}{(u^4-1)^2},
\eea 
it has four double poles at
\bea 
\mathfrak{p}=\{\pm 1,\pm i\}
\eea 
and two triple zeros at
\bea 
\mathfrak{z}=\{0,\infty\}.
\eea 
The one-form \eqref{SuperOneForm} satisfies the reality condition \eqref{real1} obviously. 
At  the  double poles we again take the Dirichlet boundary conditions \eqref{Dirichlet}. 
It was already pointed out in \cite{CY3} that there is a subtlety about the formulation of superstring coset model. In the Green-Schwarz formulation \cite{GSF}, the kinetic term of the coset Lagrangian is degenerate on the fermionic fields but the degeneracy can be resolved by introducing the $\kappa$-symmetry. So in order to construct the integrable superstring theory in this formulation from the 4D perspective, the 4D CS theory has to be modified to incorporate the $\kappa$--symmetry. This modified 4D CS theory was recently proposed in \cite{STRING}. To avoid this subtlety we will use the pure spinor formulation \cite{PureF} following the convention in \cite{CY3}. In other words, we want to construct an integrable supercoset model whose kinetic term is non-degenerate. It turns out that the corresponding Lax connection coincides with pure-spinor Lax connection \cite{SpinConnetion}. Since $\omega$ has  zeros of order $3$, in our ansatz of the Lax connection there should be the poles at the positions of the zeros with orders up to three:
\bea \label{LaxAnsatz}
&&\mathcal{L}_+=V^0_++V^1_+ u^{-3}+V^2_+ u^{-2}+V^3_+ u^{-1},\nn
&&\mathcal{L}_-=V^0_-+V^1_- u^{1}+V^2_- u^{2}+V^3_- u^{3},
\eea 
which is subjected to the reality condition $\theta(\mathcal{L})=C^{\star}(\mathcal{L})$.
Let us first take the field contents at the four boundaries to be 
\bea \label{Field}
\hat{g}|_{z=1}=g_1,\quad \hat{g}|_{z=i}=g_2,\quad \hat{g}|_{z=-1}=g_3,\quad \hat{g}|_{z=-i}=g_4,
\eea 
with the reality constraint $\Theta(g_2)=\rho^2(g_2)=g_4$.
Considering that  $\pm i$ and $\pm 1$ have the same preimage, we can set them to be connected by the $Z_4$ automorphsim $\rho$  as
\bea \label{FieldZ4}
g_2=\rho(g_1),\quad g_3=\rho^2(g_1),\quad g_4=\rho^3(g_1),\quad \rho^4=1,
\eea  
which is compatible with the reality condition and remove the overall gauge redundancy \cite{JT}. Because of the Dirichlet boundary conditions, all gauge field $A$ vanish at the boundaries so there is no local gauge redundancy to remove.
Substituting the \eqref{LaxAnsatz} and \eqref{Field} into boundary equations  gives
\bea 
&&j_{1,\pm}=V^0_\pm+V^1_\pm+V^2_\pm+V^3_\pm,\quad j_{2,\pm}=V^0_\pm+iV^1_\pm-V^2_\pm-iV^3_\pm,\nn
&&j_{3,\pm}=V^0_\pm-V^1_\pm+V^2_\pm-V^3_\pm,\quad j_{4,\pm}=V^0_\pm-iV^1_\pm-V^2_\pm+iV^3_\pm,
\eea 
where we have defined the left-invariant currents
\bea 
j_{i,\pm}=g_i^{-1}\p_{\pm}g_i.
\eea 
These equations can be solved by
\bea 
&&V^0_{\pm}=\frac{j_{1,\pm}+j_{2,\pm}+j_{3,\pm}+j_{4,\pm}}{4},\quad V^1_{\pm}=\frac{j_{1,\pm}-ij_{2,\pm}-j_{3,\pm}+ij_{4,\pm}}{4},\nn
&&V^2_{\pm}=\frac{j_{1,\pm}-j_{2,\pm}+j_{3,\pm}-j_{4,\pm}}{4},\quad V^3_{\pm}=\frac{j_{1,\pm}+ij_{2,\pm}-j_{3,\pm}-ij_{4,\pm}}{4}.
\eea 
We find that  $V_\pm^i=\theta(C^{*}V_\pm^i)$ therefore the reality condition of the Lax connection is satisfied also.
Evaluating the residues at the positions of the poles gives
\bea \label{ResSuper}
&&\text{Res}_1(\omega(u)\mathcal{L}_\pm)=\frac{1}{8}\left(\mp 3 j_{1,\pm}+(i\pm 1)j_{2,\pm}\pm j_{3,\pm}-(i\mp 1)j_{4,\pm}\right),\\
&&\text{Res}_i(\omega(u)\mathcal{L}_\pm)=\frac{1}{8}\left(\mp 3 j_{2,\pm}+(i\pm 1)j_{3,\pm}\pm j_{4,\pm}-(i\mp 1)j_{1,\pm}\right),\nn
&&\text{Res}_{-1}(\omega(u)\mathcal{L}_\pm)=\frac{1}{8}\left(\mp 3 j_{3,\pm}+(i\pm 1)j_{4,\pm}\pm j_{1,\pm}-(i\mp 1)j_{2,\pm}\right),\nn
&&\text{Res}_{-i}(\omega(u)\mathcal{L}_\pm)=\frac{1}{8}\left(\mp 3 j_{4,\pm}+(i\pm 1)j_{1,\pm}\pm j_{2,\pm}-(i\mp 1)j_{3,\pm}\right).\nonumber
\eea 
To proceed, let us consider the $\mathbb{Z}_4$ automorphism on the $\mathbb{Z}_4$ graded algebra
\bea\label{Z4} 
\rho^s(\mathfrak{g}^{(l)})=i^{sl}\mathfrak{g}^{l}, \quad l\in \{0,1,2,3\}.
\eea 
The $\mathbb{Z}_4$ transformation \eqref{FieldZ4} induces a transformation on the left-invariant currents as
\bea \label{currentZ4}
j_k=\rho^{k-1}(j_1),\quad k\in \{1,2,3,4\}.
\eea 
If we decompose the left-invariant currents into four components according to the $Z_4$-grading as
\bea \label{DecomCurrent}
j_k=j_k^{(0)}+j_{k}^{(1)}+j_k^{(2)}+j_k^{(3)},
\eea 
then the $\mathbb{Z}_4$ transformation \eqref{currentZ4} implies
\begin{align}
\label{GradedCurrent}
&j_{1}\equiv j^{(0)}+j^{(1)}+j^{(2)}+j^{(3)},\\
&j_{2}= j^{(0)}+ij^{(1)}-j^{(2)}-ij^{(3)},\nn
&j_{3}= j^{(0)}-j^{(1)}+j^{(2)}-j^{(3)},\nn
&j_{4}= j^{(0)}-ij^{(1)}-j^{(2)}+ij^{(3)}\nonumber.
\end{align}
Substituting \eqref{GradedCurrent} and \eqref{ResSuper} into the expression of the 2D action, we end up with
\bea\label{action0}
S[g]=\int \mbox{Str}(3 j_+^{(3)} j_-^{(1)}+2 j_+^{(2)} j_-^{(2)}+j_+^{(1)} j_-^{(3)}) d\sigma _+\wedge d\sigma_-,
\eea 
which coincides with the action of $AdS_5\times S^5$ superstring in the pure spinor formulation \cite{PureF}. 
\section{$\lambda$-deformation of $AdS_5\times S^5$ superstring}
\label{LambdaSuperstring}
\renewcommand{\theequation}{5.\arabic{equation}}
\setcounter{equation}{0}
The $\lambda$-deformation of $AdS_5\times S^5$ superstring has been proposed   in the Green-Schwarz formulation \cite{SuperLambda} and  in the pure spinor formulation \cite{Spin}, respectively. In this section we will use the 4D CS theory to derive it in the pure spinor formulation and put the derivation in Green-Schwarz formulation  in the Appendix.
We start by splitting  the four double poles in \eqref{SuperOneForm} into four pairs of simple poles so the one-form can be
\begin{align}
\label{equ:1.8}
\omega=\frac{u^3}{[(u-1)^2-\alpha^2][(u-i)^2+\alpha^2][(u+i)^2+\alpha^2][(u+1)^2-\alpha^2]}du ,
\end{align}
with eight simple poles  
\begin{align}
\label{equ:1.9}
\mathfrak{p}=\{1\pm\alpha,i(1\pm\alpha),-(1\pm\alpha),-i(1\pm\alpha)\}
\end{align}
and two triple zeros 
\begin{align}
\label{equ:1.10}
\mathfrak{z}=\{0,\infty\}. 
\end{align}
This 1-form \eqref{equ:1.8} satisfies \eqref{real1} when $\alpha \in \mathbb{R}$.

It is straightforward to get
\begin{align}
\label{equ:1.11}
&\operatorname{Res}_{1+\alpha}\omega=\operatorname{Res}_{i(1+\alpha)}\omega=\operatorname{Res}_{-(1+\alpha)}\omega=\operatorname{Res}_{-i(1+\alpha)}\omega\equiv K,  \notag\\
&\operatorname{Res}_{1-\alpha}\omega=\operatorname{Res}_{i(1-\alpha)}\omega=\operatorname{Res}_{-(1-\alpha)}\omega=\operatorname{Res}_{-i(1-\alpha)}\omega\equiv -K.
\end{align}
As we discussed before, the boundary conditions should be taken to be\footnote{Since $\{(x,x)\,|x\in \mathfrak{g}^\mathbb{C}\}$ is still a Lagrangian subalgebra.}
\bea
\label{equ:1.12}
&&A|_{1+\alpha}=A|_{1-\alpha}, \quad A|_{i(1+\alpha)}=A|_{i(1-\alpha)},\nn &&A|_{-(1+\alpha)}=A|_{-(1-\alpha)},\quad A|_{-i(1+\alpha)}=A|_{-i(1-\alpha)}. 
\eea
In other words we require at each pair of the simple poles the gauge fields to take values in the Lagrangian subalgebra \eqref{BoundaryLambda}. Following the argument after \eqref{BoundaryLambda} we can using the local gauge symmetry to set the fields at the boundaries to be
\begin{gather}\label{gg}
\hat{g}|_{(1+\alpha)}=g_1,\quad
\hat{g}|_{i(1+\alpha)}=g_2,\quad
\hat{g}|_{-(1+\alpha)}=g_3,\qquad
\hat{g}|_{-i(1+\alpha)}=g_4\\
\hat{g}|_{(1-\alpha)}=\hat{g}|_{i(1-\alpha)}=\hat{g}|_{-(1-\alpha)}=\hat{g}|_{-i(1-\alpha)}=1,
\end{gather}
which through \eqref{ConnectionLax} lead to 
\begin{gather}
\label{equ:1.13}
A|_{(1+\alpha)}=-\mathrm{d}g_1 g_1^{-1}+\operatorname{Ad}_{g_1}\mathcal{L}|_{(1+\alpha)},\qquad A|_{i(1+\alpha)}=-\mathrm{d}g_2 g_2^{-1}+\operatorname{Ad}_{g_2}\mathcal{L}|_{i(1+\alpha)},\\
A|_{-(1+\alpha)}=-\mathrm{d}g_3 g_3^{-1}+\operatorname{Ad}_{g_3}\mathcal{L}|_{-(1+\alpha)},\qquad A|_{-i(1+\alpha)}=-\mathrm{d}g_4 g_4^{-1}+\operatorname{Ad}_{g_4}\mathcal{L}|_{-i(1+\alpha)},\\
A|_{(1-\alpha)}=\mathcal{L}|_{(1-\alpha)},\qquad 
A|_{i(1-\alpha)}=\mathcal{L}|_{i(1-\alpha)},\\
A|_{-(1-\alpha)}=\mathcal{L}|_{-(1-\alpha)},\qquad
A|_{-i(1-\alpha)}=\mathcal{L}|_{-i(1-\alpha)}. 
\end{gather}
We can still fix $\rho^2(g_2)=g_4$ to ensure the reality condition.
Though  we can use the same ansatz \eqref{LaxAnsatz} for the Lax connection, it is more convenient to use the following  ansatz\footnote{Note that if we take the orders of the poles in the ansatz to be at most two, we will get the $AdS_5\times S^5$ model in the Green-Schwarz formulation instead of the pure spinor formulation, so as its $\lambda$-deformation, whose kinematic term is non-degenerate.} 
\begin{align}
\label{equ:1.14}
\mathcal{L}_+(u)&=U_{+0}+\frac{u}{1+\alpha}U_{+1}+\frac{u^2}{(1+\alpha)^2}U_{+2}+\frac{u^3}{(1+\alpha)^3}U_{+3},\\
\label{equ:1.14.1}
\mathcal{L}_-(u)&=\frac{(1+\alpha)^3}{u^3}U_{-3}+\frac{(1+\alpha)^2}{u^2}U_{-2}+\frac{1+\alpha}{u}U_{-1}+U_{-0},
\end{align}
where $U_{\pm0,\pm1,\pm2,\pm3}$  are regular and  take value in $\mathfrak{g}=\mathfrak{ps u}(2,2 | 4)$ so that the Lax connection is real. 
Rewriting the gauge fields in terms of the Lax connection through \eqref{ConnectionLax} and substituting into the boundary condition give a set of equations
\bea
&&-\p_\pm g_1 g_1^{-1}+\operatorname{Ad}_{g_1}\sum_{k=0}^3 U_{\pm k}=\sum_{k=0}^3\lambda^{\pm k}U_{\pm k},\\
&& -\p_\pm g_2 g_2^{-1}+\operatorname{Ad}_{g_2}\sum_{k=0}^3 i^{\pm k}U_{\pm i}=\sum_{k=0}^3(i\lambda)^{\pm k}U_{\pm k},\\
&& -\p_\pm g_3 g_3^{-1}+\operatorname{Ad}_{g_3}\sum_{k=0}^3 (-1)^{\pm k}U_{\pm i}=\sum_{k=0}^3(-\lambda)^{\pm k}U_{\pm k},\\
&& -\p_\pm g_4 g_4^{-1}+\operatorname{Ad}_{g_4}\sum_{k=0}^3 (-i)^{\pm k}U_{\pm i}=\sum_{k=0}^3(-i\lambda)^{\pm k}U_{\pm k},
\eea 
which are equivalent to
\bea \label{LambdaEqn}
&&j_{1,\pm}=\sum_{k=0}^3(1-\lambda^{\pm k}\operatorname{Ad}_{g_1}^{-1}) U_{\pm k},\quad j_{2,\pm}=\sum_{k=0}^3i^{\pm k}(1-\lambda^{\pm k}\operatorname{Ad}_{g_2}^{-1}) U_{\pm k}, \\
&&j_{3,\pm}=\sum_{k=0}^3(-1)^{\pm k}(1-\lambda^{\pm k}\operatorname{Ad}_{g_3}^{-1}) U_{\pm k},\quad j_{4,\pm}=\sum_{k=0}^3(-i)^{\pm k}(1-\lambda^{\pm k}\operatorname{Ad}_{g_4}^{-1}) U_{\pm k}. \nonumber
\eea 
Here we have defined $j_{k}\equiv g^{-1}_k\mathrm{d}g_k$, and the parameter $\lambda=(1-\alpha)/(1+\alpha)$. 

To proceed, we need to impose $\mathbb{Z}_4$ symmetry \eqref{FieldZ4} and \eqref{currentZ4}, and  then we have the decomposition of the currents \eqref{DecomCurrent} and \eqref{GradedCurrent}. Let us parametrize a group field $g$ as
\bea \label{Groupfield}
g=\operatorname{exp}\left(\sum_{k=0}^3\sum_{i_k=1}^{\operatorname{dim}(\mathfrak{g}^{(k)})}\theta_{i_k}^{(k)}T^{(k)}_{i_k}\right),
\eea 
then the $\mathbb{Z}_4$ action on the field $g$ is explicitly given by
\begin{align}
\label{Z4group}
\rho(g)=\operatorname{exp}\left(\sum_{k=0}^3\sum_{i_k=1}^{\operatorname{dim}(\mathfrak{g}^{(k)})}\theta_{i_k}^{(k)}\rho(T^{(k)}_{i_k})\right)=\operatorname{exp}\left(\sum_{k=0}^3\sum_{i_k=1}^{\operatorname{dim}(\mathfrak{g}^{(k)})}i^k \theta_{i_k}^{(k)}T^{(k)}_{i_k}\right),
\end{align}
where $T^{(k)}_{i_k}$ are the generators of the subalgebra $\mathfrak{g}^{(k)}$.  Using \eqref{FieldZ4} \eqref{Groupfield} and \eqref{Z4} and following \cite{YB2,JT} one can derive the following identity 
\begin{align}
\label{Ad}
P^{(m)} \circ \mathrm{Ad}_{g_{k}}^{-1}=\sum_{r=0}^{3} i^{(m-r)(k-1)} P^{(m)} \circ \mathrm{Ad}_{g_1}^{-1} \circ P^{(r)},
\end{align}
where $P^{(m)}$ denote projection operator onto the subalgebra $\mathfrak{g}^{(m)}$. 
For convenience, we define
\begin{gather}
\label{equ:1.26}
\mathbf{Ad}_g^{-1(p)}:=\frac{1}{4}\left(\operatorname{Ad}_{g_1}^{-1}+i^p\operatorname{Ad}_{g_2}^{-1}+i^{2p}\operatorname{Ad}_{g_3}^{-1}+i^{3p}\operatorname{Ad}_{g_3}^{-1}\right),
\end{gather}
which implies
\begin{align}
\label{equ:1.29}
P^{(m)} \circ \mathbf{Ad}_g^{-1(p)}=P^{(m)} \circ \operatorname{Ad}_{g_1}^{-1} \circ P^{(r)},
\end{align}
where $r=p+m\, \text{mod}\, 4$ and $r\in\{0,1,2,3\}$. 
Substituting \eqref{GradedCurrent} into the \eqref{LambdaEqn} we find the following rewriting of \eqref{LambdaEqn}
\bea\label{Rewriting}
(1-\mathbf{Ad}_g^{-1(0)})U_{+0}-\lambda\mathbf{Ad}_g^{-1(1)}U_{+1}-\lambda^2 \mathbf{Ad}_g^{-1(2)}U_{+2}-\lambda^3\mathbf{Ad}_g^{-1(3)}U_{+3}=&j_+^{(0)},\notag\\
-\mathbf{Ad}_g^{-1(3)}U_{+0}+(1-\lambda\mathbf{Ad}_g^{-1(0)})U_{+1}-\lambda^2 \mathbf{Ad}_g^{-1(1)}U_{+2}-\lambda^3 \mathbf{Ad}_g^{-1(2)}U_{+3}=&j_+^{(1)},\notag\\
-\mathbf{Ad}_g^{-1(2)}U_{+0}-\lambda\mathbf{Ad}_g^{-1(3)}U_{+1}+(1-\lambda^2\mathbf{Ad}_g^{-1(0)})U_{+2}-\lambda^3\mathbf{Ad}_g^{-1(1)}U_{+3}=&j_+^{(2)},\notag\\
-\mathbf{Ad}_g^{-1(1)}U_{+0}-\lambda\mathbf{Ad}_g^{-1(2)}U_{+1}-\lambda^2\mathbf{Ad}_g^{-1(3)}U_{+2}+(1-\lambda^3\mathbf{Ad}_g^{-1(0)})U_{+1}=&j_+^{(3)},\notag\\
-\lambda^{-3}\mathbf{Ad}_g^{-1(1)}U_{-3}-\lambda^{-2}\mathbf{Ad}_g^{-1(2)}U_{-2}-\lambda^{-1}\mathbf{Ad}_g^{-1(3)}U_{-1}+(1-\mathbf{Ad}_g^{-1(0)})U_{-0}=&j_-^{(0)},\notag\\
(1-\lambda^{-3}\mathbf{Ad}_g^{-1(0)})U_{-3}-\lambda^{-2}\mathbf{Ad}_g^{-1(1)}U_{-2}-\lambda^{-1}\mathbf{Ad}_g^{-1(2)}U_{-1}-\mathbf{Ad}_g^{-1(3)}U_{-0}=&j_-^{(1)},\notag\\
-\lambda^{-3}\mathbf{Ad}_g^{-1(3)}U_{-3}+(1-\lambda^{-2}\mathbf{Ad}_g^{-1(0)})U_{-2}-\lambda^{-1}\mathbf{Ad}_g^{-1(1)}U_{-1}-\mathbf{Ad}_g^{-1(2)}U_{-0}=&j_-^{(2)},\notag\\
-\lambda^{-3}\mathbf{Ad}_g^{-1(2)}U_{-3}-\lambda^{-2}\mathbf{Ad}_g^{-1(3)}U_{-2}+(1-\lambda^{-1}\mathbf{Ad}_g^{-1(0)})U_{-1}-\mathbf{Ad}_g^{-1(1)}U_{-0}=&j_-^{(3)}.\nn
\eea
After this rewriting it is easy to see that the equations \eqref{LambdaEqn} can be solved by
\begin{align}
\label{Solution}
U_+=\frac{1}{1-\operatorname{Ad}_{g_1}^{-1}\circ\Omega_+}j_+,\qquad &\Omega_+=P^{(0)}+\lambda P^{(1)}+\lambda^2 P^{(2)}+\lambda^3 P^{(3)},\\
U_-=\frac{1}{1-\operatorname{Ad}_{g_1}^{-1}\circ\Omega_-}j_-, \qquad &\Omega_-=P^{(0)}+\lambda^{-3} P^{(1)}+\lambda^{-2}P^{(2)}+\lambda^{-1}P^{(3)},
\end{align}
with
\begin{align}
\label{}
&U_{+0}=P^{(0)}U_{+}\qquad U_{+1}=P^{(1)}U_{+}\qquad U_{+2}=P^{(2)}U_{+}\qquad U_{+3}=P^{(3)}U_{+},\\
&U_{-0}=P^{(0)}U_{-}\qquad U_{-3}=P^{(1)}U_-\qquad U_{-2}=P^{(2)}U_-\qquad U_{-1}=P^{(3)}U_-.
\end{align}
Indeed $U_{\pm0,\pm1,\pm2,\pm3}$ take values in $\mathfrak{g}=\mathfrak{ps u}(2,2 | 4)$ since the currents $j_\pm$ will take values in $\mathfrak{g}=\mathfrak{ps u}(2,2 | 4)$ in the resulting 2D theory.

In the end substituting the Lax connection into the \eqref{Action} we obtain the 2D action
\bea\label{action2}
S&=&\frac{K}{2}\int \sum_{k=1}^4\left[ \text{Str}(\rho^{k-1}U_+,j_{k,-})-\text{Str}(\rho^{k-1}U_-,j_{k,+})\right]d\sigma_+\wedge d\sigma_-\nn
&=&\frac{K}{2}\int [\text{Str}(U_+,j_-)-\text{Str}(U_-,j_{,+})]d\sigma_+\wedge d\sigma_-\nn
&=&\frac{K}{2}\int [\text{Str}(j_+,j_-)-2~\text{Str}(j_+,\frac{1}{1-\operatorname{Ad}_{g_1}^{-1}\circ\Omega_-}j_-)]d\sigma_+\wedge d\sigma_-,
\eea 
where we have used $(\rho^k)^\dagger=\rho^{4-k}$ and $\Omega^{T}_+\Omega_-=\mathbf{1}$ with
\begin{align}
\label{equ:1.31}
\Omega^{T}_+=P^{(0)}+\lambda^3 P^{(1)}+\lambda^2 P^{(2)}+\lambda P^{(3)}.
\end{align}
Comparing \eqref{gg} with \eqref{Field} we find that the topological term will not be modified since the topological term only gets contributions from the pole where $\hat{g}\neq 1$. Therefore we conclude that the action \eqref{action2} we constructed in this section coincides with the one in \cite{SuperLambda} up to a prefactor.

\section{Discussion}
In this paper, we have successfully constructed the $\lambda$-deformation of the $AdS_5\times S^5$ superstring from the 4D CS theory. The same analysis is applicable for other superstring theories with $Z_4$-grading superalgebra \cite{Chen:2005uj}. It is known that the $\lambda$-deformation is Poisson-Lie-T-dual to the $\eta$-deformation \cite{PT1,PT2}. With the same one-form \eqref{equ:1.8} but choosing the other Mannin pair $(\mathfrak{g}_R,\mathfrak{d})$ as boundary conditions one should be able to derive $\eta$-deformation of the $AdS_5\times S^5$ superstring.

From the construction we realize that the discrete symmetry plays a crucial role. By orbifolding the Riemann surface we can add several copies of 2D IFTs which are related by  discrete symmetry transformations. This supplies the complement of the gluing process \cite{CY3} for constructing new IFTs. One example is the Yang-Baxter model in the  trigonometric description \cite{YB1}. We expect that  the asymmetric $\lambda$-deformation \cite{ASY} , the anisotropic $\lambda$-deformation \cite{SU2} and the generalized $\lambda$-deformation \cite{GenLambda1} can be constructed in this fashion. It may also be helpful to understand how to realize the sine-Gordon model. The sine-Gordon model can be reproduced from the affine Gaudin model by considering the Coxeter automorphism \cite{Gaudin}. Since the Gaudin model approach and the 4D CS theory approach are closely related \cite{Relation}, hence it would be interesting to figure out how the Coxeter automorphism is implemented in the 4D CS theory.

\section*{Acknowledgments}
JT and YJH would like to thank the Tohoko University for the hospitality during the 14th Kavli Asian Winter School and to thank Jue Hou, Han Liu and Jun-Bao Wu for useful discussion. The work was in part supported by NSFC Grant  No.~11335012, No.~11325522 and No. 11735001.

\appendix
\section*{Appendix: the  Green Schwarz formulation}
\label{GS}
\renewcommand{\theequation}{A.\arabic{equation}}
\setcounter{equation}{0}
In our discussion in the pure spinor formulation, we chose the ansatz \eqref{equ:1.14}\eqref{equ:1.14.1} so that the highest order of the poles of the ansatz is equal to the order of the zeros of the one-form $\omega$. If we disregard the degeneracy of the 4D action we can construct the $AdS_5\times S^5$ superstring in the Green Schwarz formulation.
To do that we simply replace the ansatz \eqref{LaxAnsatz} with	
\begin{align}
\label{equ:2.1}
&\mathcal{L}_{+}(u)=\frac{(1+\alpha)^2}{u^2}V_{-2,+}+\frac{1+\alpha}{u}V_{-1,+}+V_{0,+}+\frac{u}{1+\alpha}V_{1,+},\\
&\mathcal{L}_{-}(u)=\frac{1+\alpha}{u}V_{-1,-}+V_{0,-}+\frac{u}{1+\alpha}V_{1,-}+\frac{u^2}{(1+\alpha)^2}V_{2,-}. 
\end{align}
Then the counterpart of \eqref{Rewriting} is

\begin{align}
\label{equ:2.3}
-\lambda^2 \mathbf{Ad}_g^{-1(2)}V_{-2,+}-\lambda\mathbf{Ad}_g^{-1(3)}V_{-1,+}+(1-\mathbf{Ad}_g^{-1(0)})V_{0,+}-\frac{1}{\lambda}\mathbf{Ad}_g^{-1(1)}V_{1,+}=&j_+^{(0)},\notag\\
-\lambda^2 \mathbf{Ad}_g^{-1(1)}V_{-2,+}-\lambda\mathbf{Ad}_g^{-1(2)}V_{-1,+}-\mathbf{Ad}_g^{-1(3)}V_{0,+}+(1-\frac{1}{\lambda}\mathbf{Ad}_g^{-1(0)})V_{1,+}=&j_+^{(1)},\notag\\
(1-\lambda^2 \mathbf{Ad}_g^{-1(0)})V_{-2,+}-\lambda\mathbf{Ad}_g^{-1(1)}V_{-1,+}-\mathbf{Ad}_g^{-1(2)}V_{0,+}-\frac{1}{\lambda}\mathbf{Ad}_g^{-1(3)}V_{1,+}=&j_+^{(2)},\notag\\
-\lambda^2 \mathbf{Ad}_g^{-1(3)}V_{-2,+}+(1-\lambda\mathbf{Ad}_g^{-1(0)})V_{-1,+}-\mathbf{Ad}_g^{-1(1)}V_{0,+}-\frac{1}{\lambda}\mathbf{Ad}_g^{-1(2)}V_{1,+}=&j_+^{(3)},\notag\\
-\lambda \mathbf{Ad}_g^{-1(3)}V_{-1,-}+(1-\mathbf{Ad}_g^{-1(0)})V_{0,-}-\frac{1}{\lambda}\mathbf{Ad}_g^{-1(1)}V_{1,-}-\frac{1}{\lambda^2}\mathbf{Ad}_g^{-1(2)}V_{2,-}=&j_-^{(0)},\notag\\
-\lambda \mathbf{Ad}_g^{-1(2)}V_{-1,-}-\mathbf{Ad}_g^{-1(3)}V_{0,-}+(1-\frac{1}{\lambda}\mathbf{Ad}_g^{-1(0)})V_{1,-}-\frac{1}{\lambda^2}\mathbf{Ad}_g^{-1(1)}V_{2,-}=&j_-^{(1)},\notag\\
-\lambda \mathbf{Ad}_g^{-1(1)}V_{-1,-}-\mathbf{Ad}_g^{-1(2)}V_{0,-}-\frac{1}{\lambda}\mathbf{Ad}_g^{-1(3)}V_{1,-}+(1-\frac{1}{\lambda^2}\mathbf{Ad}_g^{-1(0)})V_{2,-}=&j_-^{(2)},\notag\\
(1-\lambda \mathbf{Ad}_g^{-1(0)})V_{-1,-}-\mathbf{Ad}_g^{-1(1)}V_{0,-}-\frac{1}{\lambda}\mathbf{Ad}_g^{-1(2)}V_{1,-}-\frac{1}{\lambda^2}\mathbf{Ad}_g^{-1(3)}V_{2,-}=&j_-^{(3)}.
\end{align}
Using \eqref{equ:1.29}, these equations can be solved to be
\begin{align}
\label{equ:2.4}
V_+=\frac{1}{1-\operatorname{Ad}_{g_1}^{-1}\circ\tilde{\Omega}_+}j_+,\qquad &\tilde{\Omega}_+=P^{(0)}+\lambda^{-1} P^{(1)}+\lambda^2 P^{(2)}+\lambda P^{(3)},\\
V_-=\frac{1}{1-\operatorname{Ad}_{g_1}^{-1}\circ\tilde{\Omega}_-}j_-, \qquad &\tilde{\Omega}_-=P^{(0)}+\lambda^{-1} P^{(1)}+\lambda^{-2}P^{(2)}+\lambda P^{(3)},  
\end{align}
which also satisfy $\tilde{\Omega}^{T}_+\tilde{\Omega}_-=\tilde{\Omega}^{T}_-\tilde{\Omega}_+=\mathbf{1}$. To be more explicit, we have
\begin{align}
\label{equ:2.5}
&V_{0,+}=P^{(0)}V_{+}\qquad V_{1,+}=P^{(1)}V_{+}\qquad V_{-2,+}=P^{(2)}V_{+}\qquad V_{-1,+}=P^{(3)}V_{+},\\
&V_{0,-}=P^{(0)}V_{-}\qquad V_{1,-}=P^{(1)}V_-\qquad V_{2,-}=P^{(2)}V_-\qquad V_{-1,-}=P^{(3)}V_-. 
\end{align}
Substituting into \eqref{Action}, the kinetic term of effective action in the Green Schwarz formulation is given by
\begin{align}
\label{equ:2.6}
S_{\text{kin}}&\propto\int\left\{\left\langle V_+,j_-\right\rangle-\left\langle V_-,j_+\right\rangle\right\}\notag\\
&\propto\int\left\{\left\langle \frac{1}{1-\operatorname{Ad}_{g_1}^{-1}\circ\tilde{\Omega}_+}j_+,j_-\right\rangle-\left\langle j_+,\frac{1}{1-\operatorname{Ad}_{g_1}^{-1}\circ\tilde{\Omega}_-}j_-\right\rangle\right\}\notag\\
&\propto\int\left\{\left\langle j_+,j_-\right\rangle-2\left\langle j_+,\frac{1}{1-\operatorname{Ad}_{g_1}^{-1}\circ\tilde{\Omega}_-}j_-\right\rangle\right\}.
\end{align}
It is same as the results in previous literature \cite{SuperLambda} up to a prefactor.


\begin{thebibliography}{99}
\bibitem{ZS}
V.~E.~Zakharov and A.~B.~Shabat,
``Integration of nonlinear equations of mathematical physics by the method of inverse scattering. II,''
Funct. Anal. Appl. \textbf{13}, 166-174 (1979)
\bibitem{YellowBook}
O.~Babelon, D.~Bernard and M.~Talon,
``Introduction to Classical Integrable Systems,''
doi:10.1017/CBO9780511535024
\bibitem{ZSfield}
D.~V.~Chudnovsky and G.~V.~Chudnovsky,
``The scheme of Zakharov-Mikhailov for two-dimensional completely integrable systems. conservation laws and backlund transformation,''
Z. Phys. C \textbf{5}, 55 (1980)
doi:10.1007/BF01546958
\bibitem{Gaudin}
B.~Vicedo,
``On integrable field theories as dihedral affine Gaudin models,''
[arXiv:1701.04856 [hep-th]].

\bibitem{Costello:2013zra}
K.~Costello,
``Supersymmetric gauge theory and the Yangian,''
[arXiv:1303.2632 [hep-th]].

\bibitem{Costello:2017dso}
K.~Costello, E.~Witten and M.~Yamazaki,
``Gauge Theory and Integrability, I,''
doi:10.4310/ICCM.2018.v6.n1.a6
[arXiv:1709.09993 [hep-th]].

\bibitem{Costello:2018gyb}
K.~Costello, E.~Witten and M.~Yamazaki,
``Gauge Theory and Integrability, II,''
doi:10.4310/ICCM.2018.v6.n1.a7
[arXiv:1802.01579 [hep-th]].

\bibitem{CY3}
	K.~Costello and M.~Yamazaki,
	``Gauge Theory And Integrability, III,''
	[arXiv:1908.02289 [hep-th]].




\bibitem{Relation}
B.~Vicedo,
``Holomorphic Chern-Simons theory and affine Gaudin models,''
[arXiv:1908.07511 [hep-th]].
\bibitem{BosonicYangBaxter}
C.~Klimcik,
``Yang-Baxter sigma models and dS/AdS T duality,''
JHEP {\bf 0212}, 051 (2002), hep-th/0210095; \\
%
C.~Klimcik,
``On integrability of the Yang-Baxter sigma-model,''
J.\ Math.\ Phys.\  {\bf 50}, 043508 (2009), arXiv:0802.3518;
%

\bibitem{DelducEta}
F.~Delduc, M.~Magro and B.~Vicedo,
``An integrable deformation of the AdS$_5$ x S$^5$ superstring action,''
Phys.\ Rev.\ Lett.\  {\bf 112}, no. 5, 051601 (2014), [arXiv:1309.5850[hep-th]];\\
``Derivation of the action and symmetries of the $q$-deformed $AdS_{5} \times S^{5}$ superstring,''
JHEP {\bf 1410}, 132 (2014), [arXiv:1406.6286[hep-th]].
\bibitem{Yoshida}
I.~Kawaguchi, T.~Matsumoto and K.~Yoshida,
``Jordanian deformations of the $AdS_5 \times S^5$ superstring,''
JHEP \textbf{04}, 153 (2014)
doi:10.1007/JHEP04(2014)153
[arXiv:1401.4855 [hep-th]].\\
T.~Matsumoto and K.~Yoshida,
``Yang–Baxter sigma models based on the CYBE,''
Nucl. Phys. B \textbf{893}, 287-304 (2015)
doi:10.1016/j.nuclphysb.2015.02.009
[arXiv:1501.03665 [hep-th]].
\bibitem{Lambda} 
K.~Sfetsos,
``Integrable interpolations: From exact CFTs to non-Abelian T-duals,''
Nucl.\ Phys.\ B {\bf 880}, 225 (2014)
[arXiv:1312.4560[hep-th]].

\bibitem{LambdaCoset} 
T.~J.~Hollowood, J.~L.~Miramontes and D.~M.~Schmidtt,
``Integrable Deformations of Strings on Symmetric Spaces,''
JHEP {\bf 1411}, 009 (2014), 
[arXiv:1407.2840[hep-th]];\\
K.~Sfetsos and D.~C.~Thompson,
``Spacetimes for $\lambda$-deformations,''
JHEP {\bf 1412}, 164 (2014)
[arXiv:1410.1886[hep-th]];
\bibitem{JT}
J.~Tian,
``Comments on $\lambda$--deformed models from 4D Chern-Simons theory,''
[arXiv:2005.14554 [hep-th]].

\bibitem{Uni}
F.~Delduc, S.~Lacroix, M.~Magro and B.~Vicedo,
``A unifying 2d action for integrable $\sigma$-models from 4d Chern-Simons theory,''
doi:10.1007/s11005-020-01268-y
[arXiv:1909.13824 [hep-th]].
\bibitem{COUPLED}
C.~Bassi and S.~Lacroix,
``Integrable deformations of coupled sigma-models,''
JHEP \textbf{20}, 059 (2020)
doi:10.1007/JHEP05(2020)059
[arXiv:1912.06157 [hep-th]].
\bibitem{HOLOLAMBDA}
D.~M.~Schmidtt,
``Holomorphic Chern-Simons theory and lambda models: PCM case,''
JHEP \textbf{04}, 060 (2020)
doi:10.1007/JHEP04(2020)060
[arXiv:1912.07569 [hep-th]].
\bibitem{YB1}
O.~Fukushima, J.~i.~Sakamoto and K.~Yoshida,
``Comments on $\eta$-deformed principal chiral model from 4D Chern-Simons theory,''
[arXiv:2003.07309 [hep-th]].
\bibitem{YB2}
O.~Fukushima, J.~i.~Sakamoto and K.~Yoshida,
``Yang-Baxter deformations of the AdS$_5\times$S$^5$ superstring from the viewpoint of 4D Chern-Simons theory,''
[arXiv:2005.04950 [hep-th]].
\bibitem{STRING}
K.~Costello and B.~Stefański,
``The Chern-Simons Origin of Superstring Integrability,''
[arXiv:2005.03064 [hep-th]].
\bibitem{GSF}
R.~R.~Metsaev and A.~A.~Tseytlin,
``Type IIB superstring action in $AdS_5 \times S^5$ background,''
Nucl. Phys. B \textbf{533}, 109-126 (1998)
doi:10.1016/S0550-3213(98)00570-7
[arXiv:hep-th/9805028 [hep-th]].
\bibitem{PureF}
N.~Berkovits,
``Super Poincare covariant quantization of the superstring,''
JHEP \textbf{04}, 018 (2000)
doi:10.1088/1126-6708/2000/04/018
[arXiv:hep-th/0001035 [hep-th]].

\bibitem{SpinConnetion}
B.~C.~Vallilo,
``Flat currents in the classical $AdS_5 \times S^5$ pure spinor superstring,''
JHEP \textbf{03}, 037 (2004)
doi:10.1088/1126-6708/2004/03/037
[arXiv:hep-th/0307018 [hep-th]].

\bibitem{SuperLambda}
T.~J.~Hollowood, J.~L.~Miramontes and D.~M.~Schmidtt,
``An Integrable Deformation of the $AdS_5 \times S^5$ Superstring,''
J. Phys. A \textbf{47}, no.49, 495402 (2014)
doi:10.1088/1751-8113/47/49/495402
[arXiv:1409.1538 [hep-th]].
\bibitem{Spin}
H.~A.~Benítez and D.~M.~Schmidtt,
``$\lambda$-deformation of the $AdS_{5}\times S^{5}$ pure spinor superstring,''
JHEP \textbf{10}, 108 (2019)
doi:10.1007/JHEP10(2019)108
[arXiv:1907.13197 [hep-th]].


\bibitem{Chen:2005uj}
B.~Chen, Y.~L.~He, P.~Zhang and X.~C.~Song,
``Flat currents of the Green-Schwarz superstrings in $AdS_5 \times S^1$ and $AdS_3 \times S^3$ backgrounds,''
Phys. Rev. D \textbf{71}, 086007 (2005)
doi:10.1103/PhysRevD.71.086007
[arXiv:hep-th/0503089 [hep-th]].
\bibitem{PT1}
B.~Vicedo,
``Deformed integrable $\sigma$-models, classical R-matrices and classical exchange algebra on Drinfel’d doubles,''
J. Phys. A \textbf{48}, no.35, 355203 (2015)
doi:10.1088/1751-8113/48/35/355203
[arXiv:1504.06303 [hep-th]].
\bibitem{PT2}
B.~Hoare and A.~A.~Tseytlin,
``On integrable deformations of superstring sigma models related to $AdS_n \times S^n$ supercosets,''
Nucl. Phys. B \textbf{897}, 448-478 (2015)
doi:10.1016/j.nuclphysb.2015.06.001
[arXiv:1504.07213 [hep-th]].
\bibitem{ASY}
S.~Driezen, A.~Sevrin and D.~C.~Thompson,
``Integrable asymmetric $\lambda$-deformations,''
JHEP \textbf{04}, 094 (2019)
doi:10.1007/JHEP04(2019)094
[arXiv:1902.04142 [hep-th]].\\
J.~Tian, J.~Hou and B.~Chen,
``Asymmetric $\lambda$-deformed cosets,''
Nucl. Phys. B \textbf{952}, 114944 (2020)
doi:10.1016/j.nuclphysb.2020.114944
[arXiv:1908.10004 [hep-th]].
\bibitem{SU2}
K.~Sfetsos and K.~Siampos,
``The anisotropic $\lambda$-deformed SU(2) model is integrable,''
Phys. Lett. B \textbf{743}, 160-165 (2015)
doi:10.1016/j.physletb.2015.02.040
[arXiv:1412.5181 [hep-th]].
\bibitem{GenLambda1}
K.~Sfetsos, K.~Siampos and D.~C.~Thompson,
``Generalised integrable $\lambda$-- and $\eta$-deformations and their relation,''
Nucl.\ Phys.\ B {\bf 899}, 489 (2015), [arXiv:1506.05784 [hep-th]];\\
Y.~Chervonyi and O.~Lunin,
``Generalized $\lambda$-deformations of AdS$_p \times$ S$^p$,''
Nucl.\ Phys.\ B {\bf 913}, 912 (2016),
[arXiv:1608.06641 [hep-th]];\\
O.~Lunin and W.~Tian,
``Analytical structure of the generalized $\lambda$-deformation,''
Nucl.\ Phys.\ B {\bf 929}, 330 (2018)
[arXiv:1711.02735 [hep-th]].


\end{thebibliography}
\end{document}